\documentclass[12pt,a4paper,english]{article}
\usepackage[T1]{fontenc}
\usepackage[latin9]{inputenc}
\usepackage{textcomp}
\usepackage{setspace}
\newcommand{\lyxaddress}[1]{
\par {\raggedright #1
\vspace{1.4em}
\noindent\par}
}

\usepackage{babel}

\begin{document}

\title{Generalized circulant matrix in Coupled Map Lattices}

\author{Mª Dolores Sotelo Herrera${}^{a}$\&Jesús San Martín${}^{a,b}$}

\date{~}

\maketitle

\lyxaddress{${}^{a}$ Departamento de Matemática Aplicada, E.U.I.T.I., Universidad
Politécnica de Madrid. Ronda de Valencia 3, 28012 Madrid Spain\\
${}^{b}$ Departamento de Física Matemática y de Fluidos, U.N.E.D.
Senda del Rey 9, 28040 Madrid Spain\\
Corresponding author: jsm@dfmf.uned.es}
\begin{abstract}
In this paper it is shown that a generalized circulant matrix underlies
every weakly Coupled Map Lattice (CML), independently of the form
of the coupling term. Therefore, this matrix will appear always perturbative
methods are used to get the analytical solutions. In fact, the inverse
of this matrix provides the analytical solution of the CML after using
first order approximation methods. This inverse matrix, of arbitrary
order, is explicitly calculated, thus providing the analytical expression
for the temporal evolution of the CML.
\end{abstract}

\section{Introduction}

Since some decades ago chaos represents an authentic revolution in
the way many problems are confronted. Its tools and applications have
spread over all sciences from mathematics to physics, passing through
chemistry, biology, economy, so on. At the beginning, chaos only studied
one-dimensional systems. The most remarkable and well-known work of
this class is the Feigenbaum's job \cite{Feigenbaum1,Feigenbaum2}
about asymptotic behaviour of unimodal function ruling equations like
\begin{equation}
x_{n+1}=f(x_{n})\label{eq:uno}\end{equation}

In spite of the amazing analytical results in one-dimensional systems,
they are not suitable to describe collective behaviour, because they
do not take into account interaction among elements of a system, so,
they can not describe many natural phenomena. One way to solve this
problem, and take into consideration interaction in a system, is by
using Coupled Map Lattices (CML) \cite{librokaneko}.

The explicit form of CML can be given by the equations

\begin{equation}
X_{i}(n+1)=(1-\alpha)f(X_{i}(n))+\frac{\alpha}{m}\sum_{i=1}^{m}f(X_{i}(n))\:\;\quad i=1,\cdots,m\label{eq:tres}\end{equation}
where $X_{i}(n)\quad i=1,\cdots,m$ represents the state of elements
in the system at time $n$, $f$ rules the individual dynamics of
any element $X_{i}(n)$ according to equation (\ref{eq:uno}) and
the interation among the individual elements is given by $\frac{\alpha}{m}\sum_{i=1}^{m}f(X_{i}(n))$;
in other cases a first neighbours interaction is chosen.

CML are used to model physical, chemical, economical and biological
systems \cite{Special1992,Special1997}. From a mathematical point
of view, they show many and interesting behaviours: travelling waves,
synchronized states, period doubling cascades, and so on. Nonetheless,
most of this behaviours have been observed numerically \cite{Palaniyandi2006,Grebogi,Czou2000,CZhou2002}
and do not have theoretical support. Furthemore, numerical results
present another problem: they are restricted by a finite and fixed
set of parameter values, specific funcions, and a finite number of
elements in the CML. Therefore, it is unknown whether or not results
are general results. On the other hand, some models need an arbitrary
number of elements in the CML to be described properly.

This problem has been recently solved in \cite{nosotros}, where the
authors get analytical results. To get the analytical solution perturbative
methods are applied and then a linear system with an arbitrary number
of equations has to be solved. An arbitrary order functional matrix
has to be inverted, at this time, the matrix is a generalized circulant
matrix, as it would be expected and we will show in section {}``conclusions
and discussion''. 

The goal of this paper is to calculate the inverse of the generalized
circulant matrix, a matrix that will underlie in every weakly coupled
map lattice. Futhermore, using the inverse matrix we will explicitly
solve the linear system which provides an analytical expression for
the time evolution of the coupled map lattice.

\section{Theoretical results}

Given that the CML\begin{equation}
X_{i}(n+1)=(1-\varepsilon\alpha)f(X_{i}(n))+\frac{\alpha\varepsilon}{p}\sum_{i=1}^{p}f(X_{i}(n))\:\;\quad i=1\cdots p\;\varepsilon\ll1\label{eq:cinco}\end{equation}
is weakly coupled, we seek a solution in the form: \[
\begin{array}{c}
X_{i}(n+j)=x_{i+j}^{*}+\varepsilon A_{i+j}\\
i=1,\cdots,p\\
j=0,\cdots,p-1\end{array}\]
where $f^{p}(x_{i}^{*})=x_{i}^{*}\quad i=1,\cdots,p$ and $f^{p^{\prime}}(x_{i}^{*})\neq1$
(bifurcation point). 

Solving the system (\ref{eq:cinco}) to order $\varepsilon$ it is
obtained:\[
-A_{i}f^{\prime}(x_{i}^{*})+A_{i+1}=\alpha x_{i}^{*}+\frac{\alpha}{p}\sum_{j=1}^{p}x_{j+1}^{*}\quad i=1,\cdots,p\]
whose matricial expression is:

{\scriptsize \begin{equation}
\left(\begin{array}{cccccc}
-f^{\prime}(x_{1}^{*}) & 1 & 0 & 0 & \cdots & 0\\
0 & -f^{\prime}(x_{2}^{*}) & 1 & 0 & \cdots & 0\\
0 & 0 & -f^{\prime}(x_{3}^{*}) & 1 & \cdots & 0\\
\vdots & \vdots & \vdots & \vdots & \ddots & \vdots\\
1 & 0 & 0 & 0 & \cdots & -f^{\prime}(x_{p}^{*})\end{array}\right)\left(\begin{array}{c}
A_{1}\\
A_{2}\\
A_{3}\\
\vdots\\
A_{p}\end{array}\right)=\alpha\left(\begin{array}{c}
-x_{2}^{*}+\frac{1}{p}\Sigma_{j=1}^{p}x_{j}^{*}\\
-x_{3}^{*}+\frac{1}{p}\Sigma_{j=1}^{p}x_{j}^{*}\\
-x_{4}^{*}+\frac{1}{p}\Sigma_{j=1}^{p}x_{j}^{*}\\
\vdots\\
-x_{1}^{*}+\frac{1}{p}\Sigma_{j=1}^{p}x_{j}^{*}\end{array}\right)\label{nueve}\end{equation}
}{\scriptsize \par}

The order $p$ matrix of the system \begin{equation}
B=\left(\begin{array}{cccccc}
-f^{\prime}(x_{1}^{*}) & 1 & 0 & 0 & \cdots & 0\\
0 & -f^{\prime}(x_{2}^{*}) & 1 & 0 & \cdots & 0\\
0 & 0 & -f^{\prime}(x_{3}^{*}) & 1 & \cdots & 0\\
\vdots & \vdots & \vdots & \vdots & \ddots & \vdots\\
1 & 0 & 0 & 0 & \cdots & -f^{\prime}(x_{p}^{*})\end{array}\right)\label{eq:alfa}\end{equation}
is not a circulant one, because $x_{1}^{*}\neq x_{2}^{*}\neq\cdots\neq x_{p}^{*}$.

The system (\ref{nueve}) has $p$ equations and $p$ unknown coefficients
whose matrix has determinant\begin{equation}
\left|B\right|=(-1)^{p}\prod_{i=1}^{p}f^{\prime}(x_{i}^{*})+(-1)^{p+1}\label{eq:aaalffa}\end{equation}

Hence, given that $\prod_{i=1}^{p}f^{\prime}(x_{i}^{*})=f^{p^{\prime}}(x_{i}^{*})\neq1$
(by hypothesis), the system is compatible and determined for every
$\alpha\neq0$. Moreover, as the independent term column is not null
because $x_{1}^{*}\neq x_{2}^{*}\neq\cdots\neq x_{p}^{*}$, it results

\[
\left(\begin{array}{c}
-x_{2}^{*}+\frac{1}{p}\sum_{j=1}^{p}x_{j}^{*}\\
-x_{3}^{*}+\frac{1}{p}\sum_{j=1}^{p}x_{j}^{*}\\
-x_{4}^{*}+\frac{1}{p}\sum_{j=1}^{p}x_{j}^{*}\\
\vdots\\
-x_{1}^{*}+\frac{1}{p}\sum_{j=1}^{p}x_{j}^{*}\end{array}\right)\neq\left(\begin{array}{c}
0\\
0\\
0\\
\vdots\\
0\end{array}\right)\]

Therefore, the solution of the system (\ref{nueve}) is different
from the trivial one, and is obtained\textbf{ }by inversion\textbf{
}as follows

\begin{equation}
\left(\begin{array}{c}
A_{1}\\
A_{2}\\
\vdots\\
A_{p}\end{array}\right)=\alpha\left(\begin{array}{ccccc}
-f^{\prime}(x_{1}^{*}) & 1 & 0 & \cdots & 0\\
0 & -f^{\prime}(x_{2}^{*}) & 1 & \cdots & 0\\
\vdots & \vdots & \vdots & \ddots & \vdots\\
1 & 0 & 0 & \cdots & -f^{\prime}(x_{p}^{*})\end{array}\right)^{-1}\left(\begin{array}{c}
-x_{2}^{*}+\frac{1}{p}\sum_{j=1}^{p}x_{j}^{*}\\
-x_{3}^{*}+\frac{1}{p}\sum_{j=1}^{p}x_{j}^{*}\\
\vdots\\
-x_{1}^{*}+\frac{1}{p}\sum_{j=1}^{p}x_{j}^{*}\end{array}\right)\label{eq:terminos}\end{equation}

The inversion of the matrix (which is not a circulant one) results
in the following\begin{equation}
\left(\begin{array}{c}
A_{1}\\
A_{2}\\
A_{3}\\
\vdots\\
A_{p}\end{array}\right)=\alpha\frac{1}{(-1)^{p+1}(1-(f^{p}(x_{1}^{*}))^{\prime})}MN\label{eq:diez}\end{equation}
where the matrix $M$ (as we will show later) is given by

{\small \begin{equation}
\left(\begin{array}{ccccc}
f^{\prime}(x_{2}^{*})\dots f^{\prime}(x_{p}^{*}) & f^{\prime}(x_{3}^{*})\dots f^{\prime}(x_{p}^{*}) & f^{\prime}(x_{4}^{*})\dots f^{\prime}(x_{p}^{*}) & \cdots & 1\\
1 & f^{\prime}(x_{3}^{*})\dots f^{\prime}(x_{p}^{*})f^{\prime}(x_{1}^{*}) & f^{\prime}(x_{4}^{*})\dots f^{\prime}(x_{p}^{*})f^{\prime}(x_{1}^{*}) & \cdots & f^{\prime}(x_{1}^{*})\\
f^{\prime}(x_{2}^{*}) & 1 & f^{\prime}(x_{4}^{*})\dots f^{\prime}(x_{p}^{*})f^{\prime}(x_{1}^{*})f^{\prime}(x_{2}^{*}) & \cdots & f^{\prime}(x_{1}^{*})f^{\prime}(x_{2}^{*})\\
\vdots & \vdots & \vdots & \ddots & \vdots\\
f^{\prime}(x_{2}^{*})\dots f^{\prime}(x_{p-1}^{*}) & f^{\prime}(x_{3}^{*})\dots f^{\prime}(x_{p-1}^{*}) & f^{\prime}(x_{4}^{*})\dots f^{\prime}(x_{p-1}^{*}) & \cdots & f^{\prime}(x_{1}^{*})f^{\prime}(x_{2}^{*})\dots f^{\prime}(x_{p-1}^{*})\end{array}\right)\label{eq:seiiis}\end{equation}
}and $N$ by

\[
\left(\begin{array}{c}
-x_{2}^{*}+\frac{1}{p}\sum_{j=1}^{p}x_{j}^{*}\\
-x_{3}^{*}+\frac{1}{p}\sum_{j=1}^{p}x_{j}^{*}\\
-x_{4}^{*}+\frac{1}{p}\sum_{j=1}^{p}x_{j}^{*}\\
\vdots\\
-x_{1}^{*}+\frac{1}{p}\sum_{j=1}^{p}x_{j}^{*}\end{array}\right)\]
After operating in (\ref{eq:diez}) it results in:

\begin{equation}
\begin{array}{c}
A_{k}=\frac{\alpha}{1-\left(f^{p}(x_{1}^{*})\right)^{\prime}}\left[\sum_{n=1}^{p-1}\left(\left(-x_{k+n}^{*}+\frac{1}{p}\sum_{j=1}^{p}x_{j}^{*}\right)\prod_{l=n}^{p-1}f^{\prime}(x_{k+l}^{*})\right)+\left(-x_{k}^{*}+\frac{1}{p}\sum_{l=1}^{p}x_{l}^{*}\right)\right]\\
k=1,\dots,p\end{array}\label{soluc}\end{equation}

Not every $A_{k}=0$, because the solution is known to be different
from the trivial one.

If some $f^{\prime}(x_{i}^{*})$ were zero then the determinant of
the matrix would be $(-1)^{p+1}$, according to equation (\ref{eq:aaalffa}),
and the inverse matrix would still exist.

We need to show that $\frac{1}{1-(f^{p}(x_{1}^{*}))^{\prime}}M$ is
the inverse of $B$ in order to prove that equation (\ref{soluc})
is the solution of system (\ref{nueve}). Inverting a funcional matrix
with arbitrary size is quite complicated, it is enough have a look
at $M$ (see eq. \ref{eq:seiiis}) to assume this conclusion. But,
if an explicit expression of the inverse matrix is known, no matter
how it is found, then the easiest way to prove that it is really the
inverse matrix is by direct multiplication. We use this technique
in the following theorem.

\paragraph*{Theorem 1}

The matrix $D=\frac{1}{(1-(f^{p}(x_{1}^{*}))^{\prime})}M$, where
{\scriptsize \[
M=\left(\begin{array}{ccccc}
f^{\prime}(x_{2}^{*})\dots f^{\prime}(x_{p}^{*}) & f^{\prime}(x_{3}^{*})\dots f^{\prime}(x_{p}^{*}) & f^{\prime}(x_{4}^{*})\dots f^{\prime}(x_{p}^{*}) & \cdots & 1\\
1 & f^{\prime}(x_{3}^{*})\dots f^{\prime}(x_{p}^{*})f^{\prime}(x_{1}^{*}) & f^{\prime}(x_{4}^{*})\dots f^{\prime}(x_{p}^{*})f^{\prime}(x_{1}^{*}) & \cdots & f^{\prime}(x_{1}^{*})\\
f^{\prime}(x_{2}^{*}) & 1 & f^{\prime}(x_{4}^{*})\dots f^{\prime}(x_{p}^{*})f^{\prime}(x_{1}^{*})f^{\prime}(x_{2}^{*}) & \cdots & f^{\prime}(x_{1}^{*})f^{\prime}(x_{2}^{*})\\
\vdots & \vdots & \vdots & \ddots & \vdots\\
f^{\prime}(x_{2}^{*})\dots f^{\prime}(x_{p-1}^{*}) & f^{\prime}(x_{3}^{*})\dots f^{\prime}(x_{p-1}^{*}) & f^{\prime}(x_{4}^{*})\dots f^{\prime}(x_{p-1}^{*}) & \cdots & f^{\prime}(x_{1}^{*})f^{\prime}(x_{2}^{*})\dots f^{\prime}(x_{p-1}^{*})\end{array}\right)\]
}is the inverse matrix of \[
B=\left(\begin{array}{cccccc}
-f^{\prime}(x_{1}^{*}) & 1 & 0 & 0 & \cdots & 0\\
0 & -f^{\prime}(x_{2}^{*}) & 1 & 0 & \cdots & 0\\
0 & 0 & -f^{\prime}(x_{3}^{*}) & 1 & \cdots & 0\\
\vdots & \vdots & \vdots & \vdots & \ddots & \vdots\\
1 & 0 & 0 & 0 & \cdots & -f^{\prime}(x_{p}^{*})\end{array}\right)\]

\paragraph*{Proof}

Let $b_{i,j}$ be the elements of the matrix $B$ and $m_{i,j}$ the
elements of the matrix $M$ , then

$b_{i,j}=\left\{ \begin{array}{cl}
-f^{\prime}(x_{i}^{*}) & \mbox{if }i=j\\
1 & \mbox{if }j=i+1,\; i\neq p\\
1 & \mbox{if }i=p,\; j=1\\
0 & \mbox{otherwise}\end{array}\right.$

$m_{i,j}=\left\{ \begin{array}{cl}
1 & \mbox{if }j=i-1,\; i\neq1\\
1 & \mbox{if }i=1,\; j=p\\
\prod_{k=1\; k\neq i}^{p}f'(x_{k}^{*}) & \mbox{if }j=i\\
m_{i-1,j}f'(x_{i-1}^{*}) & \mbox{\mbox{otherwise}}\end{array}\right.$

Let us denote by $c_{i,j}$ the elements of matrix $C$, where $C=BM$,
then it yields

\[
c_{i,j}=\sum_{k=1}^{p}b_{ik}m_{kj}=b_{i,i}m_{i,j}+b_{i,i+1}m_{i+1,j}\]
hence:
\begin{enumerate}
\item [\foreignlanguage{english}{i)}] When $j=i-1$, $i\neq1$ it results

$c_{i,i-1}=b_{ii}m_{i,i-1}+b_{i,i+1}m_{i+1,i-1}=-f^{\prime}(x_{i}^{*})+f^{\prime}(x_{i}^{*})=0$

\item [\foreignlanguage{english}{ii)}] When $i=1$, $j=p$ it results

$c_{1,p}=b_{11}m_{1,p}+b_{1,2}m_{2,p}=-f^{\prime}(x_{1}^{*})+f^{\prime}(x_{1}^{*})=0$

\item [\foreignlanguage{english}{iii)}] when $i=j$ it results

$\begin{array}{rl}
c_{i,i}= & b_{i,i}m_{i,i}+b_{i,i+1}m_{i+1,i}\\
= & -f^{\prime}(x_{i}^{*})\prod_{k=1\; k\neq i}^{p}f^{\prime}(x_{k}^{*})+1=-\prod_{k=1}^{p}f^{\prime}(x_{k}^{*})+1=-(f^{p}(x_{1}^{*}))^{\prime}+1\end{array}$

\item [\foreignlanguage{english}{iv)}] Otherwise

$c_{i,j}=b_{ii}m_{i,j}+b_{i,i+1}m_{i+1,j}=-f^{\prime}(x_{i}^{*})m_{i,j}+m_{i,j}f^{\prime}(x_{i}^{*})=0$

\end{enumerate}
Summarizing  $C=(1-(f^{p}(x_{1}^{*}))^{\prime})I$, with $I$ the
order $p$ identity matrix, and the theorem is proved.

\paragraph*{Remark.}

Observe that if, instead of having a global neighbour interaction
CML, we had any other, the generalised circulant matrix would not
change because the coupling terms lie in the intependent term column
(see equation (\ref{eq:terminos})).

\section*{Conclusions and discussion}

Most of the results in CML are numerical, despite the great efforts
to get analytical results in this field \cite{Ahmed,Chuanjun}. An
interesting theoretical result was given by Lemaitré and Chat \cite{Lemaitre}
who proved that global properties emerge from local ones in CML. 

This result is very relevant in weakly coupled map lattices because
a perturbative solution can be seeked from local behaviour. Perturbative
techniques lead to linear problems. In this paper, the theorems are
proved that allow to get the solution of the CML.

In the linearized CML given by (\ref{nueve}) one would expect to
find a circulant matrix taking into account the periodic boundary
conditions of the CML. However, the matrix of system (\ref{nueve})
must also take into account the local properties, due to the fact
that global properties emerge from them. Therefore, the matrix of
system (\ref{nueve}) is one that simultaneously includes periodic
boundary conditions and local properties, resulting in a generalized
circulant matrix. As we have pointed out, global properties emerge
from local ones, so this kind of generalized circulant matrix must
underlie every weakly CML. Its inverse, explicitly calculated in this
paper, gives the sought for analytical solutions of the CML, independently
of the form of the coupling term (see Remark after the proof of the
theorem).

\end{document}